\documentclass[epj]{svjour}
\input epsf

\newcounter{append}
\newcommand{\bc}{\begin{center}}
\newcommand{\ec}{\end{center}}
\newcommand{\be}{\begin{equation}}
\newcommand{\ee}{\end{equation}}
\newcommand{\ba}{\begin{array}}
\newcommand{\ea}{\end{array}}
\newcommand{\beqn}{\begin{eqnarray}}
\newcommand{\eeqn}{\end{eqnarray}}

\begin{document}

\title{Scaling and front dynamics in Ising quantum chains}

\author{Thierry Platini and Dragi Karevski }
\institute{Laboratoire de Physique des Mat\'eriaux, UMR CNRS No. 7556, Universit\'e Henri
Poincar\'e (Nancy 1), B.P. 239,\\ F-54506 Vand\oe uvre l\`es Nancy cedex,
France}

\date{September 15, 2005}

\abstract
{We study the relaxation dynamics of a quantum Ising chain initially prepared in a product of canonical states corresponding each to an equilibrium state of part of the chain at a given temperature. 
We focus our attention on the transverse magnetization for which a general expression is given. Explicite results are given for the completely factorized initial state, corresponding to a situation where all the spins are thermalized independently, and for the two-temperatures initial state, where part of the chain called the system is thermalized at a temperature $T_s$ and the remaining part is at a temperature $T_b$. 
}

\authorrunning{T. Platini and D. Karevski }
\titlerunning{Scaling and front dynamics in Ising quantum chains}


\maketitle

\maketitle

\section{Introduction}
Out of equilibrium properties of quantum chains have been the purpose of several recent studies (see Ref.~\cite{karebook} for a recent review). In particular, after precursor studies \cite{niem,tjio,bamc} at the end of the sixties, some renewed interest has been put on the relaxation properties of free fermionic quantum chains initially prepared in nonequilibrium states  \cite{beri,anra,sctr,bebe,igri,kare,ogat}. The influence of randomness was also considered in this context \cite{abka}.
In this contribution, we present the relaxation of the transverse magnetization of an Ising quantum chain. The system is initially prepared in a factorized state $\rho(0)=\prod_j \rho_j$ where the density matrices 
$\rho_j\propto e^{-\beta_j {\cal H}_j}$ 
are canonical states of  parts of the chain thermalized each at  a given temperature $\beta_j^{-1}$. One can look at that state as obtained from a chain initially split into several non-interacting terms: ${\cal H}_0=\sum_j {\cal H}_j$, each in contact with a specific heat bath.  
At time $t=0$, the interactions between the different parts are switched on, the heat baths removed, so that the time evolution of the system is generated quantum mechanically by the total Hamiltonian  ${\cal H}={\cal H}_0+\sum_j {\cal H}^I_{j,j+1}$, such that the system will relax from the initial factorized state $\rho(0)$ toward a new state. 

In this paper, we present the transverse magnetization profiles obtained on two different situations. The first considered is a situation where the chain is split into two parts, one at low temperature called the "system", the other at high temperature called the "bath".
So, in the system-bath case, we concentrate on the front dynamics generated by the interface between the system and the bath parts \cite{kare,anra,ogat}. The other situation  we have considered in this study is the droplet-like case, where a low temperature finite part of size $N$ is put in contact at both ends with high temperature infinite chains. In this case, we concentrate on the large time $t\gg N$ properties of the transverse magnetization, which shows a scaling behaviour. The paper is organized as follows: in the next section we present the dynamics of the Ising quantum chain solved in terms of Clifford operators \cite{kare}. Then we show how we calculate the expectation values of several physical quantities, paying a special attention to the transverse magnetization. Section~5 contains our results and we end with a short summary and discussion of them.

\section{Dynamics}
The Hamiltonian of the system under consideration is given by 
\be
{\cal H} = -\frac{1}{2} \sum_{k=1}^{L-1}\sigma_{k}^{x}\sigma_{k+1}^{x}
-\frac{h}{2}\sum_{k=1}^L\sigma_{k}^{z}
\label{h1}
\ee
where the $\sigma$s are the Pauli matrices and where $h$ is the transverse field pointing in the $z$-direction. As it is well known, this Hamiltonian can be fermionized thanks to the Jordan-Wigner transformation \cite{jowi28}:
\be
\sigma_n^x=\prod_{j=1}^{n-1}(-\sigma_j^z)\Gamma_n^1
\label{eq1}
\ee
\be
\sigma_n^y=-\prod_{j=1}^{n-1}(-\sigma_j^z)\Gamma_n^2
\label{eq2}
\ee
so that 
\be
\sigma_n^z=-i\Gamma_n^2\Gamma_n^1
\ee
where the $\Gamma$s are Clifford operators satisfying the anticommutation rules 
\be
\{\Gamma_n^i,\Gamma_m^j\}=2\delta_{n,m}\delta_{i,j}\; .
\ee
Introducing the $2L$ component operator $\Gamma$ such that
\be
\Gamma^\dagger=(\Gamma_1^1,\Gamma_1^2,\Gamma_2^1,\Gamma_2^2,
\dots,\Gamma_L^1,\Gamma_L^2)
\ee
the Hamiltonian is put in the quadratic form
\be
{\cal H}=\frac{1}{4}\Gamma^\dagger T \Gamma
\ee
where $T$ is a $2L\times2L$ matrix.
The diagonalization of the Hamiltonian can then be performed by the introduction of the diagonal Clifford operators $\gamma_q^1$ and $\gamma_q^2$, related to the lattice operators via $\Gamma_n^1=\sum_q\phi_q(n)\gamma_q^1$ and 
$\Gamma_n^2=\sum_q\psi_q(n)\gamma_q^2$ with $\phi$ and $\psi$ defined through the eigenvalue equation
\be
TV_q=\epsilon_q V_q 
\ee
with $V_q(2n-1)=\phi_q(n)$ and $V_q(2n)=-i \psi_q(n)$.
It leads to the form
\be 
{\cal H}= i\sum_q \frac{\epsilon_q}{2}\gamma_q^1\gamma_q^2\; .
\ee
The time evolution of the diagonal operators is obtained simply  by
$\gamma_{q}(t)=U_{q}^{\dagger}(t)\gamma_{q}U_{q}(t)$ with
\begin{equation}
U_{q}(t)=\exp\left(\frac{\epsilon_{q}t}{2}\gamma_{q}^{1}\gamma_{q}^{2}\right)=\cos \frac{\epsilon_{q}t}{2}
+\gamma_{q}^{1}\gamma_{q}^{2}\sin\frac{\epsilon_{q}t}{2}\; .
\end{equation}
Utilising the fact that $\{\gamma_{q}^{i},\gamma_{q'}^{j}\}=2\delta_{ij}\delta_{qq'}$, 
we obtain
\begin{equation}
\gamma_{q}^{i}(t)=\sum_{j=1}^{2}\langle \gamma_{q}^{j}|\gamma_{q}^{i}(t)\rangle\gamma_{q}^{j}
\end{equation}
where we have defined the pseudo-scalar product as
\begin{equation}
\langle C|D\rangle= \frac{1}{2}\{C^{\dagger},D\}
\end{equation}
with $\{.,.\}$ the anticommutator. The time-dependent lattice Clifford generators,
$\Gamma_{n}^{i}(t)$, can then be re-expressed in terms of the initial time operators $\Gamma$
with the help of the inverse transforms $\gamma_{q}^{1}=\sum_{k}\phi_{q}(k)\Gamma_{k}^{1}$
and $\gamma_{q}^{2}=\sum_{k}\psi_{q}(k)\Gamma_{k}^{2}$. Finally, one obtains
\begin{equation}
\Gamma_{n}^{j}(t)=\sum_{k,i}\langle \Gamma_{k}^{i}|\Gamma_{n}^{j}(t)\rangle \Gamma_{k}^{i}
\end{equation}
with components \index{Pair contractions}
\begin{eqnarray}
\langle \Gamma_{k}^{1}|\Gamma_{n}^{1}(t)\rangle &=&
 \sum_{q}\phi_{q}(k)\phi_{q}(n)\cos \epsilon_{q}t\nonumber\\
\langle \Gamma_{k}^{1}|\Gamma_{n}^{2}(t)\rangle &=&
\langle \Gamma_{n}^{2}|\Gamma_{k}^{1}(-t)\rangle =-\sum_{q}\phi_{q}(k)\psi_{q}(n)\sin \epsilon_{q}t\nonumber\\
\langle \Gamma_{k}^{2}|\Gamma_{n}^{2}(t)\rangle &=&
 \sum_{q}\psi_{q}(k)\psi_{q}(n)\cos \epsilon_{q}t\; .
\label{eqcomp}
\end{eqnarray}
One may notice  that the time evolution of the $2L$-components operator $\Gamma$ is given by a $2L$-rotation matrix $R$ with elements
\be
R_{i,j}(t)=
\langle \Gamma_{j}|\Gamma_{i}(t)\rangle 
\ee
where $\Gamma_{2i-1}\equiv \Gamma_i^1$ and $\Gamma_{2i}\equiv \Gamma_i^2$. The time evolution of $\Gamma$ is then 
\be
\Gamma(t)=R(t)\Gamma \; .
\ee

The set 
$\{\Gamma_{k}^{i}\}$ forms an orthonormal basis of a $2L$-dimensional linear 
vector space $\cal E$ with inner product defined by
$\langle.|.\rangle\equiv \frac{1}{2}\{.^{\dagger},.\}$. Hence, every vector $X\in {\cal E}$ 
has a unique expansion $X=\sum_{i,k}\langle\Gamma_{k}^{i}|X\rangle\Gamma_{k}^{i}$. 
The string expression $X_{1}X_{2}...X_{n}$, with $X_{j}\in {\cal E}$,
is a direct product vector of the space 
${\cal E}_{1}\otimes{\cal E}_{2}\otimes...\otimes{\cal E}_{n}$ which decomposition is
\begin{equation}
X_{1}X_{2}...X_{n}=\sum_{i_{1},k_{1},...,i_{n},k_{n}}
\langle\Gamma_{k_{1}}^{i_{1}}|X_{1}\rangle...
\langle\Gamma_{k_{n}}^{i_{n}}|X_{n}\rangle
\Gamma_{k_{1}}^{i_{1}}...\Gamma_{k_{n}}^{i_{n}}\; .
\label{eqstring}
\end{equation}

\section{Expectation values}
In this work, we have considered systems prepared in factorized initial states of the form
\be
\rho(0)=\prod_j\rho_j
\ee
where the initial density matrix $\rho(0)$ is given by a product of smaller density matrices $\rho_j$, corresponding to initially non-interacting subsystems.   To be more specific, we have used the canonical initial distributions
\be
\rho_j=\frac{1}{Z_j}\exp(-\beta_j {\cal H}_j)
\ee
where $\beta_j$ is the inverse temperature of the subsystem ${\cal H}_j$ which is a part of the full Hamiltonian $\cal H$. $Z_j$ is the corresponding partition function.
The expectation value of an observable $\cal O$ at time $t$ is given by
\be 
\langle {\cal O}\rangle(t) = {\rm Tr} \{O(t) \prod_j\rho_j\}
\ee
where $O(t)$ is the operator associated to the observable $\cal O$ in the Heisenberg picture.

In particular, in the following we will pay some attention on the two-temperatures problem 
\be 
\rho(0)=\rho_s\rho_b
\ee
with a small system of size $N$ initially prepared at a temperature $T_s$ putted in contact at one end with a big system of size $M$ (called the bath) at a temperature $T_b$. After the contact, the time evolution will be governed by the Hamiltonian~(\ref{h1}) with $L=N+M$. 
The expectation value takes the form
\be
\langle {\cal O}\rangle(t) = \frac{1}{Z_sZ_b}{\rm Tr} 
\{O(t) e^{-\beta_s {\cal H}_s} e^{-\beta_b {\cal H}_b}\}
\ee
with $O(t)=e^{i{\cal H}t}Oe^{-i{\cal H}t}$ and where
the total Hamiltonian is split into
\be
{\cal H}={\cal H}_s+{\cal H}_b+{\cal H}^I
\ee
with 
\be
{\cal H}_s = -\frac{1}{2} \sum_{k=1}^{N-1}\sigma_{k}^{x}\sigma_{k+1}^{x}
-\frac{h}{2}\sum_{k=1}^N\sigma_{k}^{z}
\; ,
\label{eqhs}
\ee
\be
{\cal H}_b = -\frac{1}{2} \sum_{k=N+1}^{N+M-1}\sigma_{k}^{x}\sigma_{k+1}^{x}
-\frac{h}{2}\sum_{k=N+1}^{N+M}\sigma_{k}^{z}
\; ,
\label{eqhs}
\ee
and where 
${\cal H}^I=-\frac{1}{2}\sigma_N^x\sigma_{N+1}^x$ is the initial interface interaction term. For exemple, in the situation where the system consists of only one spin, expressing the bath part in terms of its diagonal fermionic operators $\eta^{\dagger},\eta$, one has for the total Hamiltonian
\be
{\cal H}= -\frac{h}{2}\sigma_{1}^{z}
+\sum_{q=1}^{M}\varepsilon^b_{q}\eta^{\dagger}_q\eta_q
-\frac{1}{2}\sigma_1^x\sum_{q=1}^{M}\phi^b_q(1)(\eta^{\dagger}_q+\eta_q)\; ,
\ee
where $\phi_q(1)$ refers for the first bath site. 
The total Hamiltonian thus obtained is the fermionic bath version of the so-called "spin-boson" model \cite{lege,weis}.
The bath degrees of freedom act on the two-state system 
by the linear coupling  $-\frac{1}{2}\sigma_1^x\xi=-\frac{1}{2}\sigma_1^x\sum_{q=1}^{M}\phi^b_q(1)(\eta^{\dagger}_q+\eta_q)$.
The present study can be interpreted in this context.

To evaluate the expectation value $\langle {\cal O}\rangle(t)$, one first has to express the operator $O(t)$ in terms of the elementary time-independant $\Gamma_n^\nu$ Clifford operators. Then, one will have to evaluate expressions of the form
\be
\frac{1}{Z_sZ_b}{\rm Tr} \{ \Gamma_{k_{1}}^{i_{1}}...\Gamma_{k_{n}}^{i_{n}}  e^{-\beta_s {\cal H}_s} e^{-\beta_b {\cal H}_b} \}
\ee
which can be calculated explicitely noticing that the Hamiltonians ${\cal H}_s$ and ${\cal H}_b$ can be expressed in terms of the lattice operators $\Gamma_k$. 

\section{Transverse magnetization}
\subsection{System plus bath initial state}
The transverse magnetization is very simply expressed in terms of the Clifford operators, since it is a local quantity. One has in the Heisenberg picture
\be
\sigma_n^z(t)=-i \Gamma_n^2(t)\Gamma_n^1(t)\; .
\ee
The expectation value of the transverse magnetization, with initial state $\rho(0)=\rho_s\rho_b$, where the first $N$ spins of the chain are prepared in the canonical state $\rho_s$ while the $M$ remaining spins are in $\rho_b$, is 
\begin{eqnarray}
\langle \sigma_n^z\rangle(t)=-i \sum_{\nu,\nu'=1}^{2}
\sum_{k,k'=1}^{N+M}
\langle \Gamma_k^\nu |\Gamma_n^2(t)\rangle
\langle \Gamma_{k'}^{\nu'} |\Gamma_n^1(t)\rangle\nonumber\\
\times {\rm Tr}\{\rho_s\rho_b \Gamma_k^\nu\Gamma_{k'}^{\nu'}\}\; .
\end{eqnarray}
The time contractions $\langle \Gamma_k^\nu |\Gamma_n^\mu(t)\rangle$ are known and given in terms of the $\phi,\psi$ which are obtained by the diagonalization of the $T$ matrix. What remains to be done is to evaluate the trace ${\rm Tr}\{\rho_s\rho_b\Gamma_k^\nu\Gamma_{k'}^{\nu'}\}$.
One distinguishes three different situations: $\Gamma_k^\nu$ and $\Gamma_{k'}^{\nu'}$ are in the system part, $\Gamma_k^\nu$ and $\Gamma_{k'}^{\nu'}$ are in the bath part, and finally one is in the system while the other is in the bath. In the first case, noticing that 
\be
\Gamma_k^{1,2}=\Gamma_{s,k}^{1,2} \quad \forall k=1\dots N
\ee
where the $\Gamma_{s,k}$s are the Clifford operators associated to the first part of the chain (disconnected from the rest) of size $N$, one has for the trace term
\be
{\rm Tr}\{\rho_s\rho_b\Gamma_k^\nu\Gamma_{k'}^{\nu'}\}
={\rm Tr}_s\{\rho_s\Gamma_{s,k}^\nu\Gamma_{s,k'}^{\nu'}\}
\ee
The same considerations, together with the fact that
\be
\Gamma_{b,k}^{1,2}= Q_s\Gamma_{k+N}^{1,2}
\ee
where the $\Gamma_{b,k}$s are associated to the bath part and where $Q_s=\prod_{j=1}^{N}(-\sigma_j^z)$ is the charge operator of the system part, leads to the expression
\be
{\rm Tr}\{\rho_s\rho_b\Gamma_{k+N}^\nu\Gamma_{k'+N}^{\nu'}\}
={\rm Tr}_b\{\rho_b\Gamma_{b,k}^\nu\Gamma_{b,k'}^{\nu'}\}
\ee
with the indices $k$ and $k'$ running from 1 to $M$.
Finally, in the last case with one Clifford operator in each sector, it is easy to show that the trace is vanishing since it is proportional to terms of the form 
${\rm Tr}_b\{e^{-\beta_b i\frac{\epsilon_{b,q}}{2}\gamma_{b,q}^1\gamma_{b,q}^2}
\gamma_{b,q}^{\nu}\}=0$. 
Taking into account all the various contributions, one obtains for the expectation value of the transverse magnetization 
\be
\langle \sigma_n^z\rangle(t)=S_n(t)+B_n(t)
\ee
where 
\beqn
S_n(t)=\sum_{k,k'=1}^{N}\left[
\langle \Gamma_{k}^1|\Gamma_{n}^{1}(t)\rangle
\langle \Gamma_{k'}^2|\Gamma_{n}^{2}(t)\rangle \right.\nonumber\\
-
\left.\langle \Gamma_{k}^1|\Gamma_{n}^{2}(t)\rangle
\langle \Gamma_{k'}^2|\Gamma_{n}^{1}(t)\rangle
\right]({\bf S})_{k,k'}
\eeqn
and
\beqn
B_n(t)=\sum_{k,k'=1}^{M}\left[
\langle \Gamma_{k+N}^1|\Gamma_{n}^{1}(t)\rangle
\langle \Gamma_{k'+N}^2|\Gamma_{n}^{2}(t)\rangle \right.\nonumber\\
-
\left.\langle \Gamma_{k+N}^1|\Gamma_{n}^{2}(t)\rangle
\langle \Gamma_{k'+N}^2|\Gamma_{n}^{1}(t)\rangle
\right]({\bf B})_{k,k'}
\eeqn
with the canonical contractions of the system
\be
({\bf S})_{k,k'}=-\sum_{q=1}^{N}\phi_{s,q}(k)\psi_{s,q}(k') \tanh\left(\beta_s\frac{\epsilon_{s,q}}{2}\right)
\ee
and the canonical contractions of the bath
\be
({\bf B})_{k,k'}=-\sum_{q=1}^{M}\phi_{b,q}(k)\psi_{b,q}(k') \tanh\left(\beta_b\frac{\epsilon_{b,q}}{2}\right)\; .
\ee

\subsection{General expressions}
The matrices ${\bf S}$ and ${\bf B}$ contain the initial state properties. They can be rewritten more explicitely in terms of the spin operators, utilising eq.~(\ref{eq1}) and eq.~(\ref{eq2}), as
\beqn
({\bf S})_{k,k'}={\rm Tr}\{\rho_s i\Gamma_{s,k}^1\Gamma_{s,k'}^2\}
= -{\rm Tr}\{\rho_s i\sigma_k^x Q^s_{k}
Q^s_{k'}\sigma_{k'}^y
\}
\eeqn
and 
\beqn
({\bf B})_{k,k'}  ={\rm Tr}\{\rho_b i\Gamma_{b,k}^1\Gamma_{b,k'}^2\} = \qquad \qquad 
\nonumber \\
\qquad \qquad -{\rm Tr}\{\rho_b i \sigma_{N+k}^x Q^b_{k}
Q^b_{k'}\sigma_{N+k'}^y
\}
\eeqn
with $Q^s_k\equiv \prod_{j=1}^{k-1}(-\sigma_j^z)$ and $Q^b_k\equiv \prod_{j=1}^{k-1}(-\sigma_{N+j}^z)$. 
Defining the $(N+M)\times (N+M)$ matrix $\bf I$ as
\be
{\bf I}=\left(
\begin{array}{cc}
{\bf S} & 0\\
0 & {\bf B}
\end{array}
\right)\label{eqIsb}
\ee
and the time-dependent element
\be
P_{n,n'}^{k,k'}(t)=
\langle \Gamma_{k}^1|\Gamma_{n}^{1}(t)\rangle
\langle \Gamma_{k'}^2|\Gamma_{n'}^{2}(t)\rangle 
-
\langle \Gamma_{k}^1|\Gamma_{n'}^{2}(t)\rangle
\langle \Gamma_{k'}^2|\Gamma_{n}^{1}(t)\rangle\; ,
\ee
it is possible to write the expectation value of the transverse magnetization as
\be
\langle \sigma_n^z\rangle(t)=\sum_{k,k'=1}^{L=N+M}
P_{n,n}^{k,k'}(t)({\bf I})_{k,k'}\; .
\label{eqsz}
\ee 

It is easy to generalize that expression for initial product canonical states of the form $\prod_{j=1}^{K}\rho_j$
since in this case, the initial matrix $\bf I$ takes the form
\be
{\bf I}=\left(
\begin{array}{cccc}
{\bf I_1}&0&\dots&0\\
0&{\bf I}_2&&\\
\vdots&&\ddots&0\\
0&\dots&0&{\bf I}_K
\end{array}
\right)
\ee
where ${\bf I}_j$ is the initial matrix of the $j$-subsystem. The transverse magnetization is given by (\ref{eqsz}).

On the same lines, we have
\be 
\langle \sigma_n^x\sigma_{n+1}^x\rangle(t)=\sum_{k,k'=1}^{L}
P_{n+1,n}^{k,k'}(t)({\bf I})_{k,k'}\; .
\label{eqsx}
\ee 
and
\be 
\langle \sigma_n^y\sigma_{n+1}^y\rangle(t)=\sum_{k,k'=1}^{L}
P_{n,n+1}^{k,k'}(t)({\bf I})_{k,k'}
\label{eqsy}
\ee 
for the bond correlations. 

\section{Transverse magnetization profiles}
\subsection{Completely  factorized initial state}
In the completely factorized initial state
\be
\rho(0)=\rho_j^{\otimes L}
\ee
where all the spins are thermalized independently with local Hamiltonian ${\cal H}_n=-\frac{h}{2}\sigma_n^z$, the initial matrix $\bf I$ is diagonal:
\be
{\bf I}_{k,k'}=\delta_{k,k'}\langle \sigma_k^z\rangle(0)
\ee
where $\langle \sigma_k^z\rangle(0)$ is the initial transverse magnetization. In this case, the transverse magnetization at later time is given by
\be
\langle \sigma_n^z\rangle(t)=\sum_k P_{n,n}^{k,k}(t)\langle \sigma_k^z\rangle(0)\; .
\ee
In the thermodynamical limit $L\rightarrow \infty$, the contractions 
$\langle \Gamma_{k}^\nu|\Gamma_{n}^{\mu}(t)\rangle$ depend only on the difference $n-k$ so that the transverse magnetization is given by a discrete convolution product
\be
\langle \sigma_n^z\rangle(t)=\sum_k F_t(n-k)\langle \sigma_k^z\rangle(0)\; .
\ee
where the Green function $F_t(n-k)=P_{n,n}^{k,k}$.
At the critical point, $h=1$, the basic contractions are expressed in terms of Bessel functions and one can show that \cite{kare,karebook}
\be
F_t(p)=J_{2p}^2(2t)-J_{2p+1}(2t)J_{2p-1}(2t)\; .
\ee
In the continuum limit, one can show that the Green function $F_t(p)=\frac{1}{t}f(\frac{p}{t})$ with the scaling function \cite{kare}
\be
f(u)=\frac{1}{\pi }\sqrt{1-u^2} \qquad |u|\le 1
\ee
and zero otherwise.
Depending on the initial state, one will have different profiles for the transverse magnetization which can be in principle explicitely obtained by performing the above convolution product. It is interesting to notice that the time behaviour of the transverse magnetization in this case is basically the same as in the situation when the starting initial state is a pure state of the form
\be
|\Psi\rangle = |\dots \sigma_k\sigma_{k+1} \dots\rangle
\ee
where $\sigma_k$ is the value of the $z$-component spin at site $k$ \cite{igri,kare}. One can see Ref.~\cite{karebook,kare} for more details and physical examples.

\subsection{Two-temperature state}
We turn now to the more complex case where part of the chain is thermalized at inverse temperature $\beta_s$ and the rest  at inverse temperature $\beta_b$. 
The transverse magnetization is given by the general formula (\ref{eqsz}) with an initial value matrix $\bf I$ which has non-diagonal terms, taking into account that within the system part, the spins are interacting with each other and consequently have non-vanishing correlations, the same beeing true for the bath-part of the chain. 

In the following, we consider two different situations. The first is concerned with the interface behaviour, that is mainly how the jump of the transverse magnetization at the boundary between the system and the bath parts will spread out. 
The second situation  deals with the case of a finite system, of $N$ interacting spins, initially at low temperature $T_s$ in contact at both ends with infinite chains at high temperature $T_b$. In this case, the initially large transverse magnetization of the system (see the Hamiltonian~(\ref{h1})) will decay toward the small bath value as time is evolving. 

Consider the first situation. In order to avoid boundary terms,  we will take the thermodynamical limit where, both  system and bath sizes are sent to infinity. The expectation value of the transverse magnetization is given by
\be
\langle \sigma_n^z\rangle(t)=\sum_{i,k} F_t^{i}(n-k)
{\bf I}_{k,k+i}
\ee
where at the critical field value $h=1$, the Green functions are  expressed in terms of Bessel functions:
\be
F_t^i(p)=(-1)^i\left[J_{2p}(2t)J_{2(p-i)}(2t)-
J_{2p+1}(2t)J_{2(p-i)-1}(2t)\right]\; .
\ee
The relation giving the transverse magnetization decomposes into a sum of discrete convolution products
\be
\langle \sigma_n^z\rangle(t)=\sum_i(F_t^{i}\star {\bf I}^i)(n)
\ee
with Green functions $F_t^{i}$ defined above. 
In the extreme case where $T_s=0$ and $T_b=\infty$, we arrive at a further simplification since in the initial $\bf I$ matrix~(\ref{eqIsb}), the $\bf B$ matrix is vanishing and the elements of the system matrix $\bf S$ are simply $({\bf S})_{ij}=-\sum_{q=1}^\infty\phi_{s,q}(i)\psi_{s,q}(j)$.
In figure~1 we show the rescaled transverse magnetization profile obtained numerically by exact diagonalisation of the Hamiltonians ${\cal H}_s$, ${\cal H}_b$ and ${\cal H}$.
\begin{figure}[ht]
\epsfxsize=7cm
\begin{center}
\mbox{\epsfbox{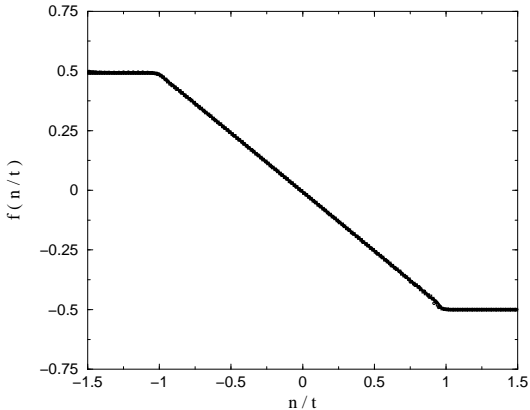}}
\end{center}
\caption{ Scaling function of the transverse magnetization obtained for times $t=30,60,90,120$ on chains of total size $L=200$.
\label{fig1}
}
\end{figure}

We have the very simple expression:
\be
\langle \sigma_n^z\rangle(t)
=\frac{1}{2}+f\left(\frac{n}{t}\right)
\label{eqsz1}
\ee
where the scaling function $f(u)$ is given by 
\be
f(u)=\left\{
\begin{array}{l}
 \frac{1}{2} \quad u\le -1 \nonumber \\
-\frac{1}{2}u \quad -1<u<1\\
-\frac{1}{2} \quad u \ge 1\nonumber 
\end{array}\right.
\label{eqsz2}
\ee
One may notice that since the excitations travel with velocity $c(h=1)=1$, one expects effects only in the causal region $-1< \ell/t<1$ where $\ell$ is the distance from the initial interface. Outside the causal region, both system and bath parts are stil equilibrated and nothing is changed since they are at equilibrium. 
The remarkable linearity of the profile is lost as 
soon as we have a departure from infinite and vanishing temperature. For finite temperature cases, the profile is  rounded. 

\begin{figure}[ht]
\epsfxsize=7cm
\begin{center}
\mbox{\epsfbox{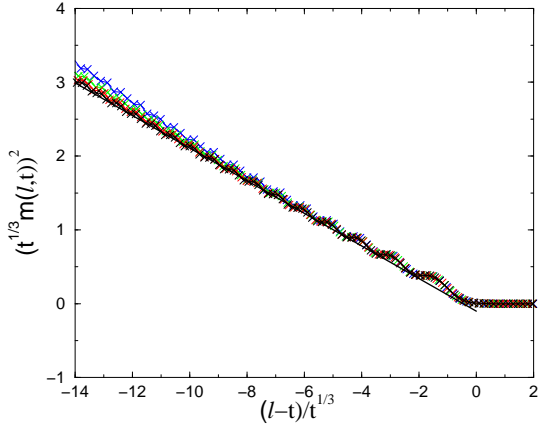}}
\end{center}
\caption{Scaling of the transverse magnetization deviation $m(\ell,t)$ obtained for times $t=80,120,160,200$ at transverse field $h=10$.
\label{fig1b}
}
\end{figure}

In the off-critical region, $h>1$, the transverse magnetization profile, in the case $T_s=0$ and $T_b=\infty$,
shows a feature that was already remarked in the XX-chain~\cite{hura} context. Indeed, as one can see on figure~2 the magnetization relaxes in quantized steps. That is, if one concentrates near the front entering into the bath part (which has a vanishing magnetization, since it is at infinite temperature), one finds a staircase like structure with constant area steps.  If one defines
\be
m(\ell,t)\equiv 
\langle \sigma_{\ell}^z\rangle(t)-\langle \sigma_{bath}^z\rangle\; ,
\ee
as it is clearly shown in figure~2, the magnetization deviation from the bath magnetization 
$\langle \sigma_{bath}^z\rangle = 0$ has a scaling behaviour of the form
\be
m(\ell,t)=t^{-1/3}g\left(\frac{\ell-t}{t^{1/3}}
\right)
\label{eqsc}
\ee
where again $\ell$ measures the distance from the initial interface, that is the interface between the system and the bath. Numerically, as seen from figure~(\ref{fig1b}), it seems that the enveloppe of the scaling function $g$ is a simple square-root, so that we have the conjecture
\be
m(\ell,t)=At^{-1/3}\sqrt{\frac{t-\ell}{t^{1/3}}} \quad \ell< t\; ,
\ee
where $A$ is a function of the field strength $h$.

The scaling form (\ref{eqsc}) implies that since the width of the steps increases as $t^{1/3}$ and their height decreases as $t^{-1/3}$, the area of the steps is indeed constant during time. Each step carries a definite magnetic moment $\mu(h)$ depending on the field strength $h$. As $h$ is increased toward infinity, the magnetic moment carried by each step reaches the value one (found numerically), which is reminiscent of the fact that in this case, an up spin from the system initial state $|\Psi\rangle=|\uparrow\uparrow\dots \uparrow\rangle$ (see the Hamiltonian~(\ref{h1})) is injected into the bath and carried by one step. 

This simple particle-like picture does not apply at the critical field value $h=1$, since in this case, the initial system state is a critical one with long range correlations.
We have seen numerically that the quantized steps picture is not present in the case $h<1$. 

Let us turn now to the finite system of size $N$, initially at vanishing temperature, in contact at both ends with infinite chains at infinite temperature. We will refer to this situation as the droplet situation.  
In figure~(3), we show the rescaled transverse magnetization profile obtained numerically at the critical field value $h=1$. 
\begin{figure}[ht]
\epsfxsize=7cm
\begin{center}
\mbox{\epsfbox{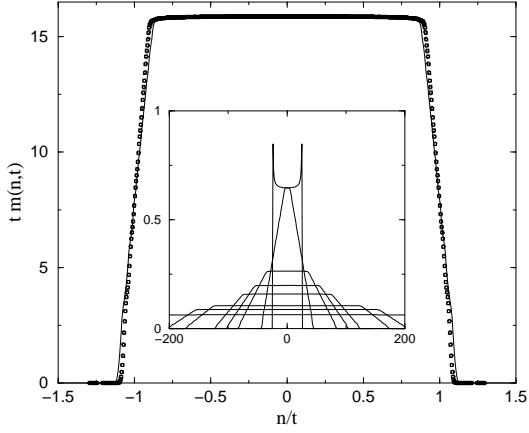}}
\end{center}
\caption{Rescaled transverse magnetization with $T_s=0$, $T_b=\infty$ and at $h=1$ for different times $t$. Inset: Transverse magnetization profiles at different times.
\label{fig2}
}
\end{figure}
We see two different regimes. 
In the first stage, for times smaller than the system size $N$, at each interface we have the phenomenology depicted previously in the infinite system case, that is a linear behaviour of the interface given by equations~(\ref{eqsz1}) and (\ref{eqsz2}). One may also remark that some transverse magnetization is lost during the initial time $t\sim N$. Indeed, since the system at $t=0$ is coupled with the bath chains, and since the total energy is conserved,  a portion of the transverse energy term, $e^z=-h/2\sigma^z$, will relax into the coupling energy term, $e^{xx}=-1/2\sigma^x\sigma^x$, leading to a net loss of transverse magnetization. For times larger than the typical time $\tau=N$, no more transverse magnetization is lost and it behaves like a conserved quantity. From the numerical profiles we see the emergence of a flat profile in the middle of the system.  The moving front between the flat finite transverse magnetization region and the zero magnetization region is linear with a slope $\sim 1/(3t)$ and has a constant width of size $N$. This leads, in the asymptotic regime $t\gg N$, to the scaling behaviour 
\be
m(n,t)\equiv \langle \sigma_{n}^z\rangle(t)
=M_{tot} \frac{1}{2t}\Pi\left(\frac{n}{2t}\right)\; ,
\label{pT0}
\ee
where $n$ is the site label measured from the middle of the system and where $M_{tot}=\sum_n m(n,t)$ is the total conserved transverse magnetization $M_{tot}=\sum_n m(n,t\gg N)$.  Numerically we found evidences for $M_{tot}=Nm^{z,sys}_{bulk}$, where $m^{z,sys}_{bulk}=2/\pi$ is the $T=0$ equilibrium system bulk transverse magnetization. 

Finally in figure~4, we show the results obtained in the droplet problem in the case where the bath part is at finite temperature.
\begin{figure}[ht]
\epsfxsize=7cm
\begin{center}
\mbox{\epsfbox{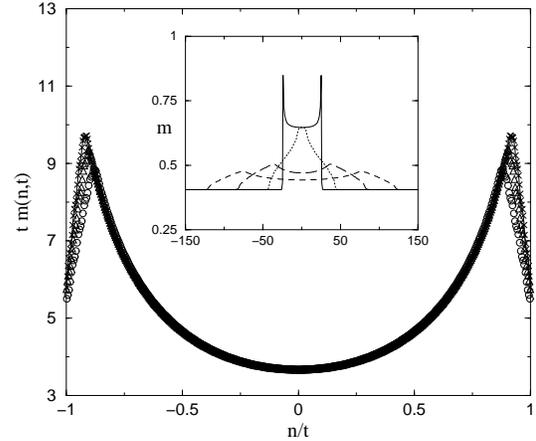}}
\end{center}
\caption{Rescaled transverse magnetization with $T_s=0$, $T_b=1$ and at $h=1$ for different times $t$. Inset: Transverse magnetization profiles at different times.
\label{fig3}
}
\end{figure}
We have more or less the same scenario as in the infinite temperature bath case, that is a velocity one propagating front, which here is not linear, of constant size $N$. In the large time limit, $t\gg N$, numerically we obtain the scaling behaviour
\be
m(n,t)=m^{z,bath}_b+\frac{N}{2t}\Pi\left(\frac{n}{2t}\right)
k\left(\frac{n}{t}\right)
\label{pT1}
\ee
with a temperature dependent function $k(u)$.
Again, the total transverse magnetization is conserved in the asymptotic limit $t\gg N$ for the same reasons as discussed previously. If one defines the excess total magnetization as $\Delta M_{tot}=\sum_n [m(n,t\gg N)-m^{z,bath}_b]$, one obtains by integration of the profile (\ref{pT1}) the relation $\Delta M_{tot}=\frac{N}{2}\int_{-1}^1 k(u) {\rm d}u$. In order to recover the scaling form~(\ref{pT0}), one has, in the limit $T_b\rightarrow\infty$, $k(u)\rightarrow 1$.
In figure~5 we show the scaling function $k(u)$ obtained for different temperatures of the bath. 
\begin{figure}[ht]
\epsfxsize=7cm
\begin{center}
\mbox{\epsfbox{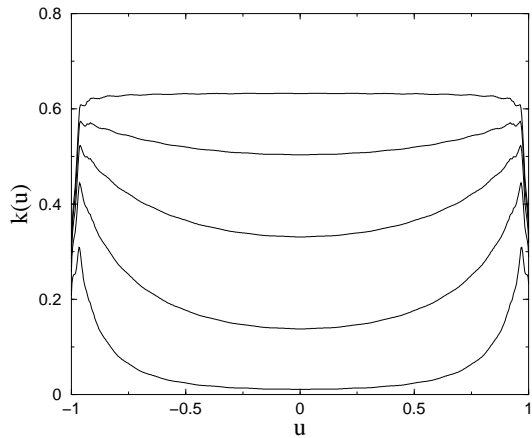}}
\end{center}
\caption{Scaling function $k(u)$ obtained for 
$T_b=\infty$, $T_b=5$,  $T_b=2$, $T_b=1$, $T_b=1/2$,
from top to bottom.
\label{fig4}
}
\end{figure}
We see clearly the flatening of the profile as the temperature increases.
  
\section{Discussion}
We have investigated the front propagation and scaling profiles of transverse magnetization inhomogeneities in the Ising quantum chain. 
The inhomogeneities were generated initially by local equilibration with several thermal baths at different temperatures. We have concentrated our study on two distinct temperature configurations: the first one beeing the interface problem, where half of the chain is at a given temperature, the other half at an other temperature. The second situation considered is one, referred as the droplet problem, where initially a small part of the infinite chain is at a different temperature from the remaining part. 
In both cases, in the asymptotic time regime the transverse magnetization exhibits a scaling form $t^{-1}M(n/t)$, where the scaling function $M$ depends on the initial temperatures and on the transverse field value $h$. At the critical field value, $h=1$, in the extreme case $T_1=0$ and $T_2=\infty$, the transverse magnetization front is linear, while whitin the droplet a flat profile emerges. Asymptotically, in the droplet situation the scaling profile is given by a characteristic function $\Pi(u)$. 
Another interesting feature of such relaxation problem is the scaling behaviour of the front itself. In particular, we have shown that in the large field situation, $h>1$, the 
front has a staircase structure with scaling  $t^{-1/3}g((\ell-t)/t^{1/3})$, meaning that the area of each step is conserved which implies a quantized relaxation of the magnetization, as already found in the XX-chain in Ref.~\cite{hura}.

\section*{Acknowledgments}
We wish to thank the Groupe de Physique Statistique for a friendly support.

\end{document}